\documentclass[twocolumn,prb]{revtex4}
\usepackage{amsfonts}
\usepackage[T1]{fontenc}
\usepackage{amsmath,amsbsy,amssymb,graphicx}
\usepackage{times}
\usepackage{amsmath}
\usepackage{amssymb}
\usepackage{graphicx}

\let\mathbf=\boldsymbol
\def\beginABC{\begin{subequations}}
\def\endABC{\end{subequations}}
\begin{document}

\title{{\Large \ Quantized Conductance and Field-Effect Topological Quantum
Transistor}\\
{\Large in Silicene Nanoribbons}}
\author{Motohiko Ezawa}
\affiliation{Department of Applied Physics, University of Tokyo, Hongo 7-3-1, 113-8656,
Japan }

\begin{abstract}
Silicene (a monolayer of silicon atoms) is a quantum spin-Hall insulator,
which undergoes a topological phase transition into other insulators by
applying external field such as electric field, photo-irradiation and
antiferromagnetic order. We investigate the electronic and transport
properties of silicene nanoribbons based on the Landauer formalism. We
propose to determine topological phase transitions by measuring the density
of states and conductance. The conductance is quantized and changes its
value when the system transforms into different phases. We show that a
silicene nanoribbon near the zero energy acts as a field-effect transistor.
This transistor is robust though it makes use of the minimum quantized
conductance since the zero-energy edge states are topologically protected.
Our findings open a new way to future topological quantum devices.
\end{abstract}

\maketitle


\address{{\normalsize Department of Applied Physics, University of Tokyo, Hongo
7-3-1, 113-8656, Japan }}

Silicene is a honeycomb structure of silicon atoms akin to graphene. Its
experimental synthesis has opened a breakthrough in the study of silicene%
\cite{GLayPRL,Takamura,Kawai}. It has two salient features absent in
graphene. One is the relatively large spin-orbit (SO) interaction, which
enables quantum spin-Hall (QSH) effects to realize\cite{LiuPRL}. The other
is its buckled structure, which enable us to apply different external fields
between the A and B sublattices such as electric field and exchange field.
As a result we can externally tune the band gap of silicene and drive
topological phase transitions from the QSH insulator to other insulators\cite%
{EzawaNJP,EzawaQAH,EzawaPhoto}. We have already shown how to make
experimental observation of phase transition points with the use of
diamagnetism\cite{Diamag} and also optical absorption\cite{EzawaOpt}.
However, these methods may not be so practical.

The most important graphene derivative is nanoribbon\cite%
{Fujita,EzawaRibbon,Brey}. The low-bias low-temperature conductance $\sigma $
of graphene nanoribbons has been shown\cite{Peres} to be quantized as%
\begin{equation}
\sigma =4\left( n+1/2\right) (e^{2}/h),\quad n=0,1,2,\cdots 
\label{AnomaSerie}
\end{equation}%
both for zigzag and armchair edges. We interpret the formula to imply the
electron-hole symmetry, the 4-fold degeneracy of each energy level
associated with the spin and valley degrees of freedom, and the conductance
quantum $e^{2}/h$ per channel. It is interesting to examine the same problem
in silicene nanoribbons because the spin-valley degeneracy is broken
according to a specific pattern in each phase.

Topological insulators are indexed\cite{Prodan,Hasan,Qi} by a set of two
topological numbers $(\mathcal{C},\mathcal{C}_{s})$, where $\mathcal{C}$ and 
$\mathcal{C}_{s}\mathcal{\ }$are\ the Chern number and the spin-Chern number
modulo $2$. There appear four types of insulators in silicene\cite%
{EzawaQAH,EzawaPhoto,Ezawa2Ferro}. They are the QSH, the quantum anomalous
Hall (QAH), the spin-polarized quantum anomalous Hall (SQAH), and the
trivial band insulators. The QSH effect is an analogue of the quantum Hall
effect for spin currents instead of charge currents. The QAH effect is the
quantum Hall effect without Landau levels, while the SQAH insulator has a
hybrid character of the QSH and QAH insulators. The prominent feature of a
topological insulator is the emergence of zero-energy edge states at
half-filling which are topologically protected against perturbation\cite%
{Hasan,Qi}.

In this paper we calculate the density of states (DOS) and the conductance
in zigzag silicene nanoribbons based on the Landauer formalism\cite{Datta},
and propose an experimentally better method to detect a topological phase
transition by way of measuring them. Our first observation is that there are
finite DOS due to the zero-energy edge states in a topological phase, while
they are absent in the trivial phase. Consequently, the topological phase
transition must be observed experimentally just by measuring the
site-resolved DOS with controlled electric field, which can be achieved by
spatially resolved the scanning tunneling microscope/scanning tunneling
spectroscopy (STM/STS).

The particularly important quantity is the conductance due to the
topologically protected zero-energy edge channels, which is the one
experimentally observable. It is interesting that the helical edge of the
QSH and the chiral edge of the QAH insulators have the same amount of
conductance. Our result is summarized as%
\begin{equation}
\begin{tabular}{||c|c|c|c|c||}
\hline\hline
topological insulator & QAH & QSH & SQAH & trivial \\ \hline
topological numbers & (2,0) & (0,1) & (1,1/2) & (0,0) \\ \hline
conductance ($\sigma $) & $2$ & $2$ & $1$ & $0$ \\ \hline\hline
\end{tabular}%
\,.  \label{BasicResult}
\end{equation}%
The conductance changes its quantized value when the system is transformed
into different phases by tuning electric field. Because of this property a
silicene nanoribbon may act as a field-effect transistor, where the
conductance is quantized and topologically protected.

\textbf{Hamiltonian:} The basic nature of silicene is described by the
tight-binding Hamiltonian\cite{KaneMele,LiuPRB}, 
\begin{equation}
H=-t\sum_{\left\langle i,j\right\rangle \alpha }c_{i\alpha }^{\dagger
}c_{j\alpha }+i\frac{\lambda _{\text{SO}}}{3\sqrt{3}}\sum_{\left\langle
\!\left\langle i,j\right\rangle \!\right\rangle \alpha \beta }\nu
_{ij}c_{i\alpha }^{\dagger }\sigma _{z}^{\alpha \beta }c_{j\beta },
\label{KaneMale}
\end{equation}%
where $c_{i\alpha }^{\dagger }$ creates an electron with spin polarization $%
\alpha $ at site $i$ in a honeycomb lattice, and $\left\langle
i,j\right\rangle /\left\langle \!\left\langle i,j\right\rangle
\!\right\rangle $ run over all the nearest/next-nearest-neighbor hopping
sites. The first term represents the usual nearest-neighbor hopping with the
transfer energy $t=1.6$eV. The second term represents the effective SO
interaction with $\lambda _{\text{SO}}=3.9$meV, and $\nu _{ij}=+1$ if the
next-nearest-neighboring hopping is anticlockwise and $\nu _{ij}=-1$ if it
is clockwise with respect to the positive $z$ axis. We have neglected the
Rashba interactions since their effects are negligibly small\cite{EzawaQAH}
in general. Here, $\sigma _{z}$ is the Pauli matrix for the the $z$
component of the spin, whose eigenvalues are $s_{z}=\pm 1$. We also use $%
s_{z}=\uparrow \downarrow $ for indices.

There are two inequivalent Brillouin zone corners, called the $K$ and $%
K^{\prime }$ points, as in graphene. The conduction and valence bands form
two conically shaped valleys or cones at these points. We assign the valley
index $\mu =\pm $ to electrons at the $K$ or $K^{\prime }$ point, and call
them the $K_{\eta }$ points as well. We also use $\eta =K,K^{\prime }$ for
indices. The low-energy effective Hamiltonian is given by the massive Dirac
theory around the $K_{\eta }$ point. The Hamiltonian (\ref{KaneMale}) yields%
\begin{equation}
H_{\eta }^{0}=\hbar v_{\text{F}}\left( \eta k_{x}\tau _{x}+k_{y}\tau
_{y}\right) +\lambda _{\text{SO}}\eta \tau _{z}\sigma _{z},
\label{DiracHamil}
\end{equation}%
where $v_{\text{F}}=\frac{\sqrt{3}}{2}at$ is the Fermi velocity with the
lattice constant $a=3.86$\AA . Here, $\tau _{z}$ is the Pauli matrix for the 
$z$ component of the pseudospin representing the $A$ and $B$ sublattices,
whose eigenvalues are $t_{z}=\pm 1$.

\begin{figure}[t]
\centerline{\includegraphics[width=0.5\textwidth]{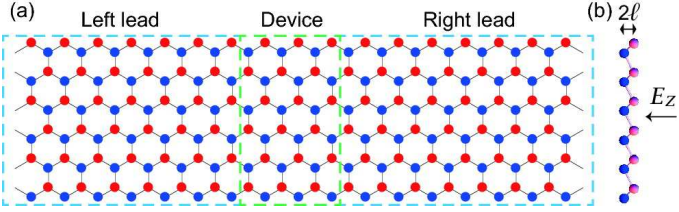}}
\caption{(Color online) Silicene nanoribbon in electric field. (a) It is
decomposed into the device, right lead and left lead parts. The width is
taken to be $W=5$. (b) Silicene consists of the $A$-sublattice and the $B$%
-sublatice with layer separation $2\ell $. The energy of the $A$ sites (red
disc) is lower than the one of the $B$ sites (blue disc) in electric field $%
E_{z}$. The edge mode is localized along the $A$ sites ($B$ sites) of the up
(down) outmost edge of a nonoribbon. See Fig.\protect\ref{FigESili} for the
site-resolved DOS of the edge modes.}
\label{FigDevice}
\end{figure}

A great merit of silicene is that we can introduce various potential terms
into the Hamiltonian by making advantages of its bucked structure. Eight
commuting terms are possible in the Dirac Hamiltonian (\ref{DiracHamil}),%
\begin{equation}
H_{pqr}=\lambda _{pqr}\eta ^{p}(\sigma _{z})^{q}(\tau _{z})^{r},
\label{pqrTerm}
\end{equation}%
where $p,q,r=0$ or $1$. Each term has different symmetry properties. The
coefficient of $\tau _{z}$ is the Dirac mass, to which four terms
contribute. They are $H_{pq1}$. First, $H_{111}$ is nothing but the SO
coupling term with $\lambda _{111}=\lambda _{\text{SO}}$. Second, $H_{001}$
is the staggered sublattice potential term\cite{EzawaNJP} with $\lambda
_{001}=\ell E_{z}$, where $2\ell $ is the separation between the $A$ and $B$
sublattices and $E_{z}$ is the external electric field. Third, $H_{101}$ is
the Haldane term\cite{Haldane}, where we set $\lambda _{101}=\lambda_{\Omega
}$ by introducing photo-irradiation\cite{EzawaPhoto,Kitagawa} with strength $%
\lambda _{\Omega }$. Finally, $H_{011}$ is the the staggered exchange term%
\cite{Ezawa2Ferro}, where we set $\lambda _{011}=\Delta M=M_{A}-M_{B}$ by
introducing the exchange fields $M_{A(B)}$ to the $A(B)$-sublattice.

We may write down the tight-binding term that yields the potential term $%
H_{pq1}$. The additional terms are\cite{EzawaNJP,EzawaQAH,Ezawa2Ferro}%
\begin{align}
\Delta H& =i\frac{\lambda _{\Omega }}{3\sqrt{3}}\sum_{\left\langle
\!\left\langle i,j\right\rangle \!\right\rangle \alpha \beta }\nu
_{ij}c_{i\alpha }^{\dagger }c_{j\beta }  \notag \\
& -\ell E_{z}\sum_{i\alpha }t_{z}^{i}c_{i\alpha }^{\dagger }c_{i\alpha
}+\Delta M\sum_{i\alpha }t_{z}^{i}c_{i\alpha }^{\dagger }\sigma
_{z}c_{i\alpha },  \label{BasicHamil}
\end{align}%
where $t_{z}^{i}=\pm 1$ for $i=A,B$. The Dirac Hamiltonian is%
\begin{equation}
H_{\eta }=H_{\eta }^{0}-\ell E_{z}\tau _{z}+\eta \lambda _{\Omega }\tau
_{z}+\Delta M\sigma _{z}\tau _{z}.  \label{TotalDirac}
\end{equation}%
The spin-valley dependent Dirac mass is given by%
\begin{equation}
\Delta _{s_{z}}^{\eta }=\eta s_{z}\lambda _{\text{SO}}-\ell E_{z}+\eta
\lambda _{\Omega }+s_{z}\Delta M.  \label{DiracMass}
\end{equation}%
It may be positive, negative or zero. The silicene system is described by
the Hamiltonian $H+\Delta H$.

\begin{figure}[t]
\centerline{\includegraphics[width=0.5\textwidth]{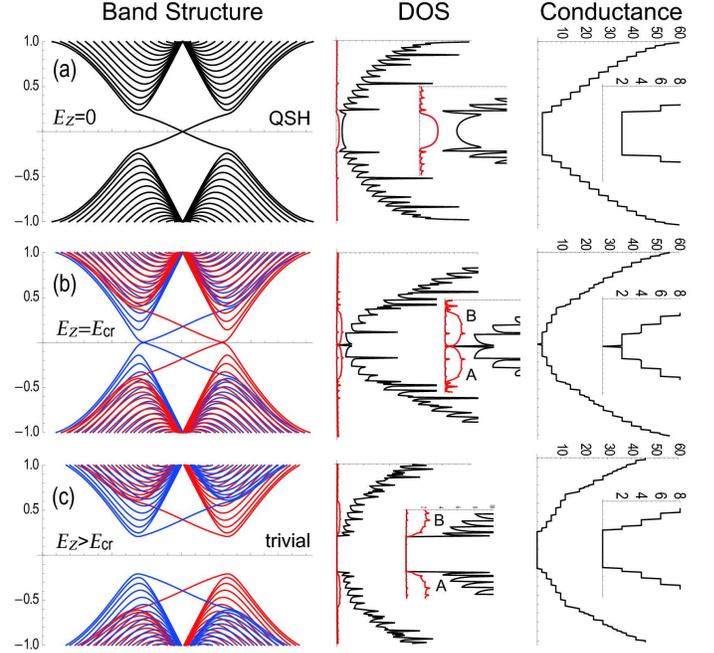}}
\caption{(Color online) Band structure, DOS and conductance of zigzag
silicene nanoribbons for (a) the QSH insulator phase, (b) the metalic phase
at the phase transition point, and (c) the trivial insulator phase. These
phases are obtained by applying electric field $E_{z}$. The phase transition
occurs at $E_{z}=E_{\text{cr}}$. The number of bands is $2W+2$ in the
nanoribbon with width $W$. Here, the width is taken to be $W=31$, and only a
part of bands are shown. The band gap is degenerate (nongenerate) with
respect to the up (red) and down (blue) spins at $E_{z}=0$ ($E_{z}>0$). Van
Hove singularities emerge in the DOS at the points where the band dispersion
is flat. The site-resolved DOS of the up-spin state at the outmost $A$ and $%
B $ sites of a nanoribbon are shown by red curves in the insets. There are
finite DOS for the zero-energy edge states in the QSH insulator. The
conductance is quantized by unit of $e^{2}/h$. }
\label{FigESili}
\end{figure}

\textbf{Topological phases:} Any insulating state is characterized by a set
of two topological quantum numbers $(\mathcal{C},\mathcal{C}_{s})$. Provided
the spin $s_{z}$ is a good quantum number, they are given by $\mathcal{C}=%
\mathcal{C}_{\uparrow }^{K}+\mathcal{C}_{\uparrow }^{K^{\prime }}+\mathcal{C}%
_{\downarrow }^{K}+\mathcal{C}_{\downarrow }^{K^{\prime }}$ and $\mathcal{C}%
_{s}=\frac{1}{2}(\mathcal{C}_{\uparrow }^{K}+\mathcal{C}_{\uparrow
}^{K^{\prime }}-\mathcal{C}_{\downarrow }^{K}-\mathcal{C}_{\downarrow
}^{K^{\prime }})$, where $\mathcal{C}_{s_{z}}^{\eta }$ is the summation of
the Berry curvature in the momentum space over all occupied states of
electrons with spin $s_{z}$ in the Dirac valley $K_{\eta }$, and calculated%
\cite{Diamag} as $\mathcal{C}_{s_{z}}^{\eta }={\frac{\eta }{2}}$sgn$(\Delta
_{s_{z}}^{\eta })$. All possible topological insulators are determined by
the three parameters $E_{z}$, $\lambda _{\Omega }$ and $\Delta M$ with
respect to $\lambda _{\text{SO}}$. Possible sets of topological numbers are $%
(0,0),(2,0),(0,\frac{1}{2}),(1,\frac{1}{2})$ up to the sign $\pm $. They are
the trivial, QAH, QSH, SQAH insulators, respectively. Note that there are
two-types of trivial band insulators, which are the charge-density-wave
(CDW) type insulator and the antiferromagnetic (AF) insulator\cite%
{Ezawa2Ferro}.

A topological phase transition occurs when the band gap closes, or $\Delta
_{s_{z}}^{\eta }=0$. Let us review\cite{EzawaNJP} the topological phase
transition along the $E_{z}$ axis, where the Dirac mass is given by $\Delta
_{s_{z}}^{\eta }=\eta s_{z}\lambda _{\text{SO}}-\ell E_{z}$. The condition $%
\Delta _{s_{z}}^{\eta }=0$ implies $E_{z}=\pm E_{\text{cr}}$ with $E_{\text{%
cr}}=\lambda _{\text{SO}}/\ell $. It follows that $(\mathcal{C},\mathcal{C}%
_{s})=(0,0)$ for $|E_{z}|<E_{\text{cr}}$ and $(0,\frac{1}{2})$ for $%
|E_{z}|>E_{\text{cr}}$. We have given the band structures at $E_{z}=0$, $E_{%
\text{cr}}$ and $2E_{\text{cr}}$ in Fig.\ref{FigESili}. It is the
characteristic feature known as the bulk-edge correspondence that
zero-energy edge modes emerge in topological insulators with $(\mathcal{C},%
\mathcal{C}_{s})\neq (0,0)$ and that these zero-energy edge modes are
topologically protected against perturbation\cite{Hasan,Qi}.

\begin{figure}[t]
\centerline{\includegraphics[width=0.5\textwidth]{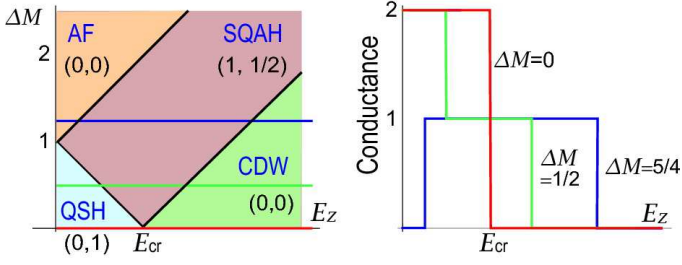}}
\caption{(Color online) (a) Phase diagram in the $(E_{z},\Delta {M})$ plane
and (b) conductance as a function of $E_{z}$ at fixed values of $\Delta {M}$%
. It takes a constant in one phase, and changes its value across the phase
boundary. The unit is $\protect\lambda _{\text{SO}}$ for $\Delta M$, and $%
e^{2}/h$ for the conductance.}
\label{FigEdM}
\end{figure}

\begin{figure}[t]
\centerline{\includegraphics[width=0.5\textwidth]{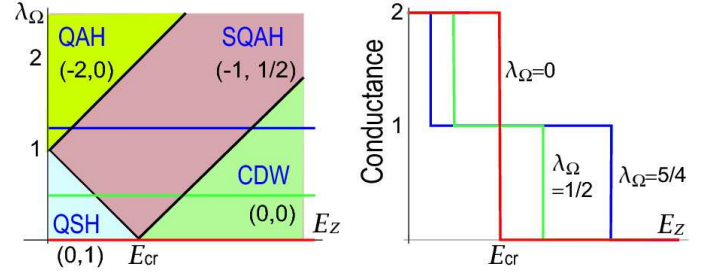}}
\caption{(Color online) (a) Phase diagram in the $(E_{z},\protect\lambda %
_{\Omega })$ plane and (b) conductance as a function of $E_{z}$ at fixed
values of $\protect\lambda _{\Omega }$. It takes a constant in one phase,
and changes its value across the phase boundary. The unit is $\protect%
\lambda _{\text{SO}}$ for $\protect\lambda _{\Omega }$, and $e^{2}/h$ for
the conductance.}
\label{FigEP}
\end{figure}

We may easily construct the phase diagrams by solving $\Delta _{s_{z}}^{\eta
}=0$. Those in the $(E_{z},\Delta M)$ space and the $(E_{z},\lambda _{\Omega
})$ space are given in Figs.\ref{FigEdM} and \ref{FigEP} together with the
pair of topological charges $(\mathcal{C},\mathcal{C}_{s})$, respectively.
All possible topological insulators are found in these phase diagrams. The
band structures at typical states are found in Figs.\ref{FigESili} and \ref%
{FigQAH-SQAH}, where the breakdown of the spin-valley symmetry is manifest.
We proceed to characterize each phase by its characteristic pattern of the
DOS and the conductance of a zigzag nanoribbon.

\textbf{DOS and conductance: }The natural framework for transport
calculations in nanoscopic devices is the Landauer formalism\cite{Datta}. We
consider a zigzag silicene nanoribbon divided into three regions [Fig.\ref%
{FigDevice}]: the device region, the left lead and the right lead. The size
of the device region is actually irrelevant due to the ballistic transport
property.

In terms of single-particle Green's functions, the low-bias conductance $%
\sigma (E)$ at the Fermi energy $E$ is given by\cite{Datta} 
\begin{equation}
\sigma (E)=(e^{2}/h)\text{Tr}[\Gamma _{\text{L}}(E)G_{\text{D}}^{\dag
}(E)\Gamma _{\text{R}}(E)G_{\text{D}}(E)],
\end{equation}%
where $\Gamma _{\text{R(L)}}(E)=i[\Sigma _{\text{R(L)}}(E)-\Sigma _{\text{%
R(L)}}^{\dag }(E)]$ with the self-energies $\Sigma _{\text{L}}(E)$ and $%
\Sigma _{\text{R}}(E)$, and%
\begin{equation}
G_{\text{D}}(E)=[E-H_{\text{D}}-\Sigma _{\text{L}}(E)-\Sigma _{\text{R}%
}(E)]^{-1},  \label{StepA}
\end{equation}%
with the Hamiltonian $H_{\text{D}}$ for the device region. The self-energy $%
\Sigma _{\text{L(R)}}(E)$ describes the effect of the electrode on the
electronic structure of the device, whose real part results in a shift of
the device levels whereas the imaginary part provides a life time. It is to
be calculated numerically\cite{Sancho,Rojas,Nikolic,Li}.

The total density of states (DOS) reads%
\begin{equation}
\rho (E)=-\pi ^{-1}\text{Im}\text{Tr}G_{\text{D}}(E),
\end{equation}%
while the partial density of states at $i$ site reads%
\begin{equation}
\rho _{i}(E)=-\pi ^{-1}\text{Im}[G_{\text{D}}(E)_{ii}],
\end{equation}%
in terms of the Green function $G_{\text{D}}(E)$ of the device.

\begin{figure}[t]
\centerline{\includegraphics[width=0.5\textwidth]{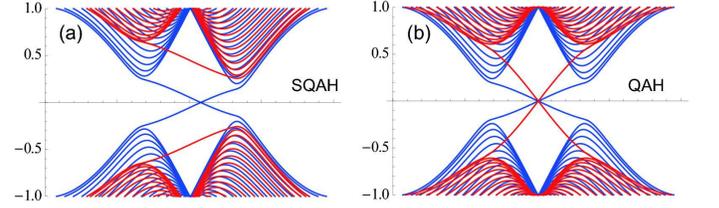}}
\caption{(Color online) Band structures of the SQAH and QAH insulators.
Up-spin (down-spin) states are illustrated in red (blue). There are two
channels in the QAH insulator but only one channel in the SQAH insulator
that contribute to the conductance at half-filling.}
\label{FigQAH-SQAH}
\end{figure}

We have calculated the DOS $\rho (E)$ and the conductance $\sigma (E)$ of a
nanoribbon as functions of the Fermi energy $E$, which is controlled by
doping. We give the results at electric field $E_{z}=0$, $E_{\text{cr}}$ and 
$2E_{\text{cr}}$ in Fig.\ref{FigESili}. A van Hove singularity occurs in the
DOS at the point where the band dispersion is flat. As $E$ increases beyond
the point, the Fermi level crosses a new band. A new channel opens and
contributes to the conductance by $e^{2}/h$ for each spin and valley. It is
clearly observed that the edge channel connects the tips of the Dirac cones
with the same spin at the $K$ and $K^{\prime }$ points.

We have also plotted the site-resolved DOS $\rho _{i}(E)$ of the up-spin
states at the outmost $A$ and $B$ sites of a nanoribbon by red curves in the
insets [Fig.\ref{FigESili}]. They represent degenerate zero-energy states at 
$E_{z}=0$. As we have explained in Fig.\ref{FigDevice}, the energy of the $A$
and $B$ sites become different for $E_{z}\neq 0$. It results in the downward
(upward) shift of $\rho _{A(B)}(E)$ along the edge as $E_{z}$ increases.
They are separated completely, and zero-energy states disappear for $%
E_{z}>E_{\text{cr}}$.

The zero-energy edge channel of a topological insulator is particularly
important at half-filling, because it is topologically protected. We have
calculated the conductance at half-filling by increasing the external field $%
E_{z}$ in the $(E_{z},\Delta M)$ space and the $(E_{z},\lambda _{\Omega })$
space. First, we increase $E_{z}$ from $E_{z}=0$ at $\Delta M=0,\frac{1}{2}%
\lambda _{\text{SO}},\frac{5}{4}\lambda _{\text{SO}}$ in the $(E_{z},\Delta
M)$ space [Fig.\ref{FigEdM}]. Initially the conductance reads $\sigma
=2e^{2}/h$ for $\Delta M=0$, where the system is in the QSH phase. It reads $%
\sigma =e^{2}/h$, when the system is in the SQAH phase. Its band structure
is shown in Fig.\ref{FigQAH-SQAH}(a), where the zero-energy edge states
account for the conductance $\sigma =e^{2}/h$. The conduction becomes zero
as $E_{z}$ increases and the system becomes the trivial AF insulator.

We confirm these observations by investigating the same problem in the $%
(E_{z},\lambda _{\Omega })$ space [Fig.\ref{FigEP}]. The system is in the
QAH phase for $\lambda _{\Omega }>\lambda _{\text{SO}}$ at $E_{z}=0$, where $%
\sigma =2e^{2}/h$. We illustrate the band structure of the QSH insulator in
Fig.\ref{FigQAH-SQAH}(b), where the zero-energy edge states account for the
conductance $\sigma =2e^{2}/h$. The edge channel is helical (chiral) in the
QSH (QAH) phase, but both of them transport the same amount of electric
charges when the current is fed. We summarize the conductance in each
topological insulator as in (\ref{BasicResult}).

\textbf{Field-effect topological quantum transistor: }The conductance is
quantized in silicene nanoribbons. The simplest system is provided by
applying electric field only\cite{EzawaNJP}, where quantized conductance changes from $2$
to $0$ at the critical electric field $E_{\text{cr}}$. This means the system
acts as a transistor where "on" state can be switched off to "off" state by
applying electric field. This transistor is "quantum" since the conductance
is quantized, which is highly contrasted with the ordinal transistor, where
the conductance is not quantized. Furthermore the conductance is
topologically protected because the zero-energy edge state is topologically
protected. Namely the conductance is robust against impurities due to its
topological stability. Consequently we may call it a field-effect
topological quantum transistor. This is the most energy-saving device since
it utilizes the minimum conductance.

We are able to design a three-digit quantum transistor by attaching
antiferromagnet. Namely the conductance changes in three steps 2, 1, 0 with
increasing electric field when $\left\vert \Delta M\right\vert <\lambda _{%
\text{SO}}$. When $\left\vert \Delta M\right\vert >\lambda _{\text{SO}}$,
the conductance changes in three step 0, 1, 0. It acts as a three-step
transistor, where the system is first in the "off" state, then become "on"
state and finally become "off" state with the increase of electric field.

\textbf{Conclusions:} We have analyzed the DOS and the conductance in
silicene nanoribbons. There are finite DOS due to the zero-energy edge
states in the topological phase, while they disappear in the trivial phase
[Fig.\ref{FigESili}]. Local DOS measurement is a direct evidence of the
existence of the edge states, which can be achieved by spatially resolved
STM/STS. This must be the most efficient way to make an experimental
observation of a topological phase transition. Furthermore we can determine
the band gap by measuring the DOS. A precise measurement is possible owing
to the van-Hove singularities present at the tips of the conduction and
valence bands [Fig.\ref{FigESili}].

Conductance measurement is also a direct method to observe a topological
phase transition. We have proposed a field-effect topological quantum
transistor with the use of the zero-energy edge state of a silicene
nanoribbon. This could be a basic component of future topological quantum
devices.

In passing we address the problem how narrow the nanoribbon can be. The
penetration depth of the zero-energy edge state has already been shown to be
as short as the atomic scale\cite{Arm} in zigzag nanoribbons. Hence we may
use a quite narrow nanoribbon to detect and make use of the topological
properties of silicene.

\label{SecConclusion}

I am very much grateful to N. Nagaosa and B. K. Nikoli\'{c} for many helpful
discussions on the subject. This work was supported in part by Grants-in-Aid
for Scientific Research from the Ministry of Education, Science, Sports and
Culture No. 22740196.

\end{document}